\documentclass[conference]{IEEEtran}
\IEEEoverridecommandlockouts
\usepackage{cite}
\usepackage{amsmath,amssymb,amsfonts}
\usepackage{algorithmic}
\usepackage{multirow}
\usepackage{graphicx}
\usepackage{float}
\usepackage{balance}
\usepackage{colortbl}
\usepackage{textcomp}
\usepackage{float}
\usepackage{cite}
\usepackage{amsmath,amssymb,amsfonts}
\usepackage{algorithmic}
\usepackage{graphicx}
\usepackage{textcomp}
\usepackage{xcolor}
 \usepackage[hidelinks]{hyperref}
\usepackage{balance}
\usepackage{multirow}
\usepackage{tabularx}

\def\BibTeX{{\rm B\kern-.05em{\sc i\kern-.025em b}\kern-.08em
    T\kern-.1667em\lower.7ex\hbox{E}\kern-.125emX}}
    
\begin{document}


\title{\textit{DBJoules}: An Energy Measurement Tool for Database Management Systems}



\author{

\IEEEauthorblockN{Hemasri Sai Lella}
\IEEEauthorblockA{\textit{Research in Intelligent Systems and Human Analytics Lab} \\
Department of Computer Science and Engineering\\
Indian Institute of Technology Tirupati, India\\
cs20b020@iittp.ac.in}

\IEEEauthorblockN{Kurra Manasa}
\IEEEauthorblockA{\textit{Research in Intelligent Systems and Human Analytics Lab} \\
Department of Computer Science and Engineering\\
Indian Institute of Technology Tirupati, India\\
cs20b019@iittp.ac.in}
\and
\IEEEauthorblockN{Rajrupa Chattaraj}
\IEEEauthorblockA{\textit{Research in Intelligent Systems and Human Analytics Lab} \\
Department of Computer Science and Engineering\\
Indian Institute of Technology Tirupati, India\\
cs22s504@iittp.ac.in}

\IEEEauthorblockN{Sridhar Chimalakonda}
\IEEEauthorblockA{\textit{Research in Intelligent Systems and Human Analytics Lab} \\
Department of Computer Science and Engineering\\
Indian Institute of Technology Tirupati, India\\
ch@iittp.ac.in}}






\maketitle
\begin{abstract}

In the rapidly evolving landscape of modern data-driven technologies, software relies on large datasets and constant data center operations using various database systems to support computation-intensive tasks.  As energy consumption in software systems becomes a growing concern, selecting the right database from energy-efficiency perspective is also critical. To address this, we introduce \textbf{\textit{DBJoules}}, a tool that measures the energy consumption of activities in database systems. \textit{DBJoules} supports energy measurement of CRUD operations for four popular databases. Through evaluations on two widely-used datasets, we identify disparities of 7\% to 38\% in the energy consumption of these databases. Hence, the goal is to raise developer awareness about the effect of running queries in different databases from an energy consumption perspective, enabling them to select appropriate database for sustainable usage. The tool's demonstration is available at \url{https://youtu.be/D1MTZum0jok} and related artifacts at \url{https://rishalab.github.io/DBJoules/}.


\end{abstract}




\maketitle

\section{Introduction}
\label{sec: Intro}
In the current era, data is proliferating at an unprecedented pace. Recent estimates indicate that data equivalent to 0.33 zettabytes (around 10\textsuperscript{12} GB) is generated on a daily basis\footnote{\url{https://explodingtopics.com/blog/data-generated-per-day\#}}. Effectively managing this enormous volume of data, in addition to the existing data, presents a significant challenge. To support this growth, a multitude of data centers, operating 24/7, have been established. However, these data centers consume a significant amount of energy. To put this into perspective, global electricity consumption by data centers was estimated to range between 240 and 340 terawatt-hours in 2022\footnote{\url{https://www.iea.org/energy-system/buildings/data-centres-and-data-transmission-networks\#}}, and is rapidly growing.

As data continues to grow over time, its efficient management through databases becomes increasingly imperative. Databases play a fundamental role in the software, functioning as repositories for the storage and administration of diverse data types. They are an integral part throughout different stages of software development, from initial application creation to deployment and continuous maintenance, ensuring data integrity, security, and accessibility. As data centers become more prevalent across various sectors and industries, it also becomes essential to examine the core of these data centers, i.e., databases from an energy-consumption perspective\cite{hintemann2019energy}. 

While the primary focus in the field of databases has traditionally been on achieving faster processing and scalability \cite{khan2023sql}, only a few studies in the literature have delved into query optimization \cite{guo2017green} and energy-efficient query processing \cite{tu2014system}. However, the energy consumption associated with basic database queries, constituting the majority of operations that a database performs, has largely been overlooked. Therefore, our first step is towards measuring the energy consumption of these queries in various databases.

\par

\begin{figure*}
\centering
\includegraphics[scale=0.285]{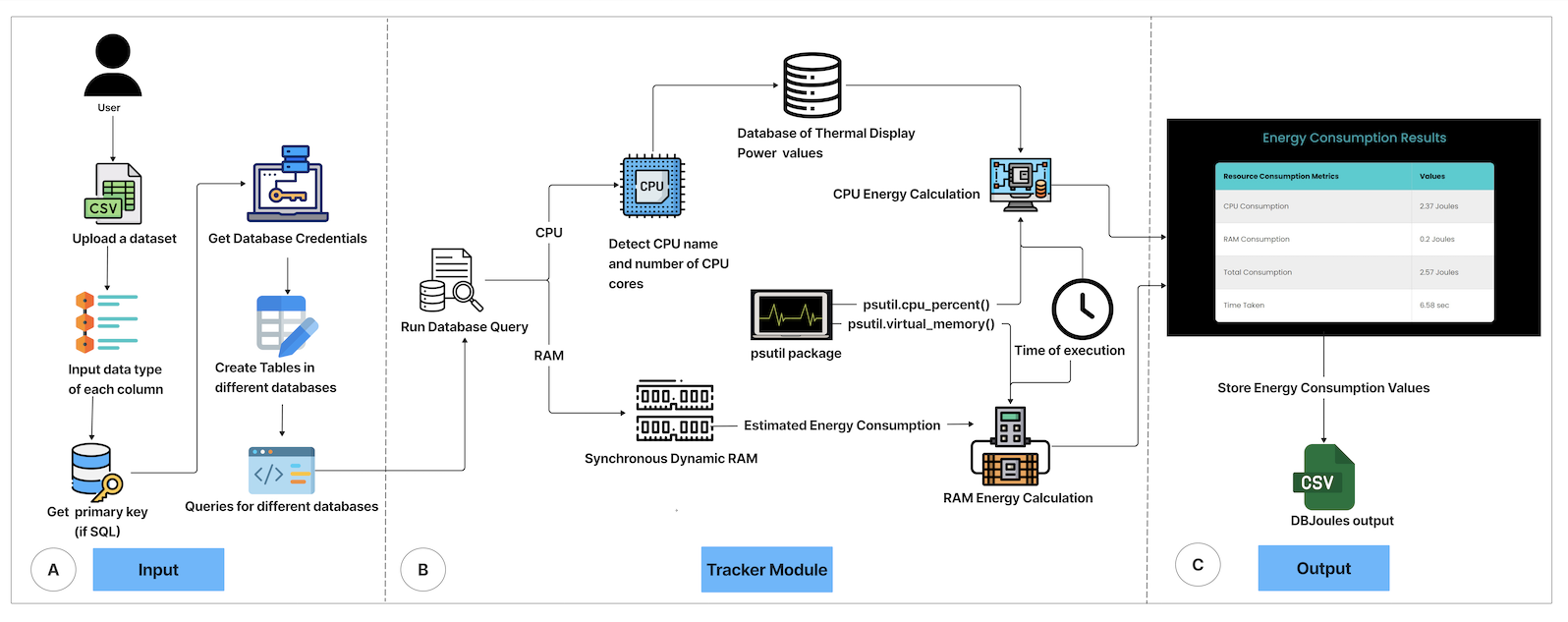}
\caption{Workflow Diagram of \textit{DBJoules}: A describes the input functionalities of \textit{DBJoules}, B describes the internal workflow of \textit{Tracker Module}, C describes the output derived from \textit{DBjoules}.}
\label{fig:architecture}
\end{figure*}


In this context, we propose \textbf{\textit{DBJoules}}, a tool designed to measure the energy consumption of database queries across four databases, i.e., MySQL, PostgreSQL, MongoDB, and Couchbase. We measure the energy by gathering the information about CPU and RAM usage, by utilizing \textit{psutil} package\footnote{\url{https://github.com/giampaolo/psutil}}. \textit{DBJoules} is evaluated by empirically examining the energy consumption of basic queries on two popular datasets stored in above mentioned databases. Based on the values obtained by DBJoules, we found that for specific queries, energy consumption may vary by up to 38\% among various databases.



\section{Related Work}
In recent years, the research community has made significant efforts to address the energy consumption of various software components. This includes programming languages \cite{couto2017towards} and data structures \cite{pereira2016influence}. Energy measurement tools such as EcoML \cite{igescuecoml}, Codecarbon\footnote{\url{https://github.com/mlco2/codecarbon}}, jRAPL\footnote{\url{https://github.com/kliu20/jRAPL}}, pyJoules\footnote{\url{https://pypi.org/project/pyJoules/}} and RJoules \cite{chattaraj2023rjoules} have emerged with the aim of assisting developers in building more energy-efficient software.


A few empirical studies in the literature \cite{couto2017towards, georgiou2022green} have identified potential areas for optimization and improvements in energy efficiency of software. For instance, Verdecchia et al. \cite{verdecchia2022data} revealed that simple modifications to datasets alone could lead to a significant reduction in energy consumption, up to 92.16\%, in machine learning algorithms. Similarly, another study found that the choice of dataframe library during the data pre-processing stage could impact energy consumption by up to 202 times \cite{shanbhag2023exploratory}. Building on this understanding, our goal is to explore the energy utilization of databases. Therefore, \textit{DBJoules} is our effort to measure the energy consumed during database querying, aiming to raise awareness among developers about the energy implications of selecting a particular database for their software.



\section{Methodology}
In this section, we will detail the design decisions that were made and describe the development process followed for \textit{DBJoules}.


\textit{DBJoules} measures the energy consumption of database queries by utilizing the \textit{psutil} (process and system utilities) library to extract CPU utilization and memory allocation associated with databases. \textit{psutil} is a cross-platform Python library that performs system monitoring and profiling while calculating process resources. It provides essential information about active processes and system usage, including CPU, memory, disks, network, and sensors. Multiple studies \cite{henderson2020towards, budennyy2022eco2ai} and software projects\footnote{\url{https://github.com/google/grr}} have demonstrated the reliability of this library for measuring energy consumption.

\textit{DBJoules} supports four databases based on their widespread popularity and extensive usage\footnote{\url{https://survey.stackoverflow.co/2023/\#section-most-popular-technologies-databases}}. These include two SQL databases, namely MySQL and PostgreSQL, and two NoSQL databases, MongoDB and Couchbase. The tool currently provides support for four fundamental database queries: SELECT, INSERT, DELETE, and UPDATE. \textit{DBJoules} offers two primary functionalities: it can measure the energy consumption of an individual query in any of the four database systems, and it can compare the energy consumed by a query executed in two or more database systems.

\textit{DBJoules}, developed in Python, relies on various Python packages for establishing connections with databases such as:
\begin{itemize}
     \item \textbf{mysql-connector-python} : This module facilitates Python programs in accessing and manipulating MySQL databases through API calls. 
     \item \textbf{psycopg2} : Psycopg is the most popular PostgreSQL database adapter for the Python. It offers complete implementation of the Python DB API 2.0 specification\footnote{\url{https://peps.python.org/pep-0249/}} and ensures thread safety.
    \item \textbf{pymongo} : This package serves as a native Python driver for MongoDB, providing tools for seamless interaction with MongoDB databases from Python.  
    \item \textbf{couchbase} : This module allows Python applications to access a Couchbase cluster efficiently.
\end{itemize}

\textit{DBJoules} workflow, illustrated in Figure \ref{fig:architecture}, begins with the user uploading a dataset in a `\texttt{.csv}' file. The tool initiates connections with the databases using the connection packages mentioned earlier. It then identifies the dataset's columns and prompts the user to specify data types and select primary key columns in case of SQL databases. The user must provide database credentials, including usernames, passwords, and database names for each database. After these steps, tables are created in all four database systems, as shown in Figure \ref{fig:architecture}A. Next, the user writes input queries for all four databases. \textit{DBJoules} calculates the energy consumption of each database using its `\textit{tracker module},' as depicted in Figure \ref{fig:architecture}B.


The tracker module performs several key tasks during its operation. First, it identifies the CPU model and the number of processors on the host machine, which are used to determine the thermal design power (TDP) values. These TDP values tell us about average power consumption specific to a CPU model and are obtained from a dataset compiled by Budennyy et al. \cite{budennyy2022eco2ai}, covering 3279 distinct processors from Intel and AMD. If the exact CPU model name cannot be found, the module employs pattern matching to identify the closest match; otherwise, it assigns a constant value of 100, as followed by Maevsky et al. \cite{maevsky2017evaluating}. 

Next, the module gathers CPU utilization percentages using the \texttt{psutil.cpu\_percent()} function and divides this value by the number of CPU cores to obtain the average CPU load. Using the extracted information, the module calculates CPU energy consumption, as per the equation \ref{eqn:cpu} below:
\begin{equation}
    \label{eqn:cpu}
    \boldsymbol{E_{CPU} = TDP \times W_{CPU} \times t}
\end{equation}
where, $E_{\text{CPU}}$ represents the total energy consumption of the CPU during query execution, $\text{W}_{\text{CPU}}$ is the CPU utilization percentage and $\text{t}$ is the CPU loading time.

Additionally, the module calculates energy consumption associated with RAM, recognizing that database operations are significantly influenced by data read or write activities. The RAM's power consumption is directly proportional to the power allocated by the currently running process. This is measured using \texttt{psutil.virtual\_memory()}, and the RAM energy consumption is determined using equation \ref{eqn:RAM} proposed by \cite{maevsky2017evaluating}:
\begin{equation}
    \label{eqn:RAM}
    \boldsymbol{E_{RAM} = 0.375 \times M_{RAM} \times t}
\end{equation}
where, $E_{\text{RAM}}$ is the RAM energy consumption, $\text{M}_{\text{RAM}}$ is the allocated memory in DB, as measured by psutil and $\text{t}$ is loading.
0.375 W/GB is the estimated specific energy consumption of DDR3 and DDR4 modules \cite{maevsky2017evaluating}. Finally, the tracker module outputs these computed values in a CSV file (Figure \ref{fig:architecture}C).

\begin{table*}[]
\caption{Mean values for energy consumption of four databases on Intel and AMD processor systems in Joules.}
\label{tab:my-table}
\centering
\begin{tabular}{|cc|llllll|llllll|}
\hline
\rowcolor[HTML]{FFFFFF} 
\multicolumn{2}{|l|}{\cellcolor[HTML]{FFFFFF}}                                                                                                 & \multicolumn{6}{l|}{\cellcolor[HTML]{FFFFFF}\textbf{Netflix Userbase Dataset}}                                                                                                                                                                                                                                                   & \multicolumn{6}{l|}{\cellcolor[HTML]{FFFFFF}\textbf{SMS Spam Collection Dataset}}                                                                                                                                                                                                                                                \\ \hline
\rowcolor[HTML]{FFFFFF} 
\multicolumn{1}{|l|}{\cellcolor[HTML]{FFFFFF}}                                    & \cellcolor[HTML]{FFFFFF}                                   & \multicolumn{3}{l|}{\cellcolor[HTML]{FFFFFF}\textbf{System1}}                                                                                                                         & \multicolumn{3}{l|}{\cellcolor[HTML]{FFFFFF}\textbf{System2}}                                                                            & \multicolumn{3}{l|}{\cellcolor[HTML]{FFFFFF}\textbf{System1}}                                                                                                                         & \multicolumn{3}{l|}{\cellcolor[HTML]{FFFFFF}\textbf{System2}}                                                                            \\ \cline{3-14} 
\rowcolor[HTML]{FFFFFF} 
\multicolumn{1}{|l|}{\multirow{-2}{*}{\cellcolor[HTML]{FFFFFF}\textbf{Database}}} & \multirow{-2}{*}{\cellcolor[HTML]{FFFFFF}\textbf{Queries}} & \multicolumn{1}{l|}{\cellcolor[HTML]{FFFFFF}\textbf{CPU}}  & \multicolumn{1}{l|}{\cellcolor[HTML]{FFFFFF}\textbf{RAM}}  & \multicolumn{1}{l|}{\cellcolor[HTML]{FFFFFF}\textbf{Total}} & \multicolumn{1}{l|}{\cellcolor[HTML]{FFFFFF}\textbf{CPU}}  & \multicolumn{1}{l|}{\cellcolor[HTML]{FFFFFF}\textbf{RAM}}  & \textbf{Total} & \multicolumn{1}{l|}{\cellcolor[HTML]{FFFFFF}\textbf{CPU}}  & \multicolumn{1}{l|}{\cellcolor[HTML]{FFFFFF}\textbf{RAM}}  & \multicolumn{1}{l|}{\cellcolor[HTML]{FFFFFF}\textbf{Total}} & \multicolumn{1}{l|}{\cellcolor[HTML]{FFFFFF}\textbf{CPU}}  & \multicolumn{1}{l|}{\cellcolor[HTML]{FFFFFF}\textbf{RAM}}  & \textbf{Total} \\ \hline
\rowcolor[HTML]{EFEFEF} 
\multicolumn{1}{|l|}{\cellcolor[HTML]{EFEFEF}}                                    & SELECT                                                     & \multicolumn{1}{l|}{\cellcolor[HTML]{EFEFEF}1.61}          & \multicolumn{1}{l|}{\cellcolor[HTML]{EFEFEF}0.18}          & \multicolumn{1}{l|}{\cellcolor[HTML]{EFEFEF}1.79}           & \multicolumn{1}{l|}{\cellcolor[HTML]{EFEFEF}\textbf{0.71}} & \multicolumn{1}{l|}{\cellcolor[HTML]{EFEFEF}\textbf{0.14}} & \textbf{0.85}  & \multicolumn{1}{l|}{\cellcolor[HTML]{EFEFEF}2.24}          & \multicolumn{1}{l|}{\cellcolor[HTML]{EFEFEF}0.19}          & \multicolumn{1}{l|}{\cellcolor[HTML]{EFEFEF}2.43}           & \multicolumn{1}{l|}{\cellcolor[HTML]{EFEFEF}0.59}          & \multicolumn{1}{l|}{\cellcolor[HTML]{EFEFEF}0.10}          & 0.69           \\ \cline{2-14} 
\rowcolor[HTML]{EFEFEF} 
\multicolumn{1}{|l|}{\cellcolor[HTML]{EFEFEF}}                                    & INSERT                                                     & \multicolumn{1}{l|}{\cellcolor[HTML]{EFEFEF}1.58}          & \multicolumn{1}{l|}{\cellcolor[HTML]{EFEFEF}0.19}          & \multicolumn{1}{l|}{\cellcolor[HTML]{EFEFEF}1.77}           & \multicolumn{1}{l|}{\cellcolor[HTML]{EFEFEF}0.74}          & \multicolumn{1}{l|}{\cellcolor[HTML]{EFEFEF}0.15}          & 0.89           & \multicolumn{1}{l|}{\cellcolor[HTML]{EFEFEF}1.67}          & \multicolumn{1}{l|}{\cellcolor[HTML]{EFEFEF}0.19}          & \multicolumn{1}{l|}{\cellcolor[HTML]{EFEFEF}1.86}           & \multicolumn{1}{l|}{\cellcolor[HTML]{EFEFEF}0.72}          & \multicolumn{1}{l|}{\cellcolor[HTML]{EFEFEF}0.10}          & 0.82           \\ \cline{2-14} 
\rowcolor[HTML]{EFEFEF} 
\multicolumn{1}{|l|}{\cellcolor[HTML]{EFEFEF}}                                    & UPDATE                                                     & \multicolumn{1}{l|}{\cellcolor[HTML]{EFEFEF}\textbf{1.43}} & \multicolumn{1}{l|}{\cellcolor[HTML]{EFEFEF}\textbf{0.18}} & \multicolumn{1}{l|}{\cellcolor[HTML]{EFEFEF}\textbf{1.61}}  & \multicolumn{1}{l|}{\cellcolor[HTML]{EFEFEF}0.73}          & \multicolumn{1}{l|}{\cellcolor[HTML]{EFEFEF}0.15}          & 0.88           & \multicolumn{1}{l|}{\cellcolor[HTML]{EFEFEF}1.89}          & \multicolumn{1}{l|}{\cellcolor[HTML]{EFEFEF}0.18}          & \multicolumn{1}{l|}{\cellcolor[HTML]{EFEFEF}2.07}           & \multicolumn{1}{l|}{\cellcolor[HTML]{EFEFEF}0.76}          & \multicolumn{1}{l|}{\cellcolor[HTML]{EFEFEF}0.09}          & 0.85           \\ \cline{2-14} 
\rowcolor[HTML]{EFEFEF} 
\multicolumn{1}{|l|}{\multirow{-4}{*}{\cellcolor[HTML]{EFEFEF}MySQL}}             & DELETE                                                     & \multicolumn{1}{l|}{\cellcolor[HTML]{EFEFEF}2.04}          & \multicolumn{1}{l|}{\cellcolor[HTML]{EFEFEF}0.19}          & \multicolumn{1}{l|}{\cellcolor[HTML]{EFEFEF}2.23}           & \multicolumn{1}{l|}{\cellcolor[HTML]{EFEFEF}0.73}          & \multicolumn{1}{l|}{\cellcolor[HTML]{EFEFEF}0.14}          & 0.87           & \multicolumn{1}{l|}{\cellcolor[HTML]{EFEFEF}\textbf{1.6}}  & \multicolumn{1}{l|}{\cellcolor[HTML]{EFEFEF}\textbf{0.18}} & \multicolumn{1}{l|}{\cellcolor[HTML]{EFEFEF}\textbf{1.78}}  & \multicolumn{1}{l|}{\cellcolor[HTML]{EFEFEF}\textbf{0.52}} & \multicolumn{1}{l|}{\cellcolor[HTML]{EFEFEF}\textbf{0.08}} & \textbf{0.60}  \\ \hline
\rowcolor[HTML]{FFFFFF} 
\multicolumn{1}{|l|}{\cellcolor[HTML]{FFFFFF}}                                    & SELECT                                                     & \multicolumn{1}{l|}{\cellcolor[HTML]{FFFFFF}1.36}          & \multicolumn{1}{l|}{\cellcolor[HTML]{FFFFFF}0.19}          & \multicolumn{1}{l|}{\cellcolor[HTML]{FFFFFF}1.55}           & \multicolumn{1}{l|}{\cellcolor[HTML]{FFFFFF}0.95}          & \multicolumn{1}{l|}{\cellcolor[HTML]{FFFFFF}0.15}          & 1.10           & \multicolumn{1}{l|}{\cellcolor[HTML]{FFFFFF}1.77}          & \multicolumn{1}{l|}{\cellcolor[HTML]{FFFFFF}0.18}          & \multicolumn{1}{l|}{\cellcolor[HTML]{FFFFFF}1.95}           & \multicolumn{1}{l|}{\cellcolor[HTML]{FFFFFF}0.56}          & \multicolumn{1}{l|}{\cellcolor[HTML]{FFFFFF}0.14}          & 0.70           \\ \cline{2-14} 
\rowcolor[HTML]{FFFFFF} 
\multicolumn{1}{|l|}{\cellcolor[HTML]{FFFFFF}}                                    & INSERT                                                     & \multicolumn{1}{l|}{\cellcolor[HTML]{FFFFFF}1.39}          & \multicolumn{1}{l|}{\cellcolor[HTML]{FFFFFF}0.19}          & \multicolumn{1}{l|}{\cellcolor[HTML]{FFFFFF}1.58}           & \multicolumn{1}{l|}{\cellcolor[HTML]{FFFFFF}0.67}          & \multicolumn{1}{l|}{\cellcolor[HTML]{FFFFFF}0.14}          & 0.81           & \multicolumn{1}{l|}{\cellcolor[HTML]{FFFFFF}2.06}          & \multicolumn{1}{l|}{\cellcolor[HTML]{FFFFFF}0.18}          & \multicolumn{1}{l|}{\cellcolor[HTML]{FFFFFF}2.24}           & \multicolumn{1}{l|}{\cellcolor[HTML]{FFFFFF}0.64}          & \multicolumn{1}{l|}{\cellcolor[HTML]{FFFFFF}0.13}          & 0.77           \\ \cline{2-14} 
\rowcolor[HTML]{FFFFFF} 
\multicolumn{1}{|l|}{\cellcolor[HTML]{FFFFFF}}                                    & UPDATE                                                     & \multicolumn{1}{l|}{\cellcolor[HTML]{FFFFFF}\textbf{1.28}} & \multicolumn{1}{l|}{\cellcolor[HTML]{FFFFFF}\textbf{0.18}} & \multicolumn{1}{l|}{\cellcolor[HTML]{FFFFFF}\textbf{1.46}}  & \multicolumn{1}{l|}{\cellcolor[HTML]{FFFFFF}\textbf{0.55}} & \multicolumn{1}{l|}{\cellcolor[HTML]{FFFFFF}\textbf{0.14}} & \textbf{0.69}  & \multicolumn{1}{l|}{\cellcolor[HTML]{FFFFFF}2.09}          & \multicolumn{1}{l|}{\cellcolor[HTML]{FFFFFF}0.18}          & \multicolumn{1}{l|}{\cellcolor[HTML]{FFFFFF}2.27}           & \multicolumn{1}{l|}{\cellcolor[HTML]{FFFFFF}\textbf{0.52}} & \multicolumn{1}{l|}{\cellcolor[HTML]{FFFFFF}\textbf{0.13}} & \textbf{0.65}  \\ \cline{2-14} 
\rowcolor[HTML]{FFFFFF} 
\multicolumn{1}{|l|}{\multirow{-4}{*}{\cellcolor[HTML]{FFFFFF}MongoDB}}           & DELETE                                                     & \multicolumn{1}{l|}{\cellcolor[HTML]{FFFFFF}1.34}          & \multicolumn{1}{l|}{\cellcolor[HTML]{FFFFFF}0.19}          & \multicolumn{1}{l|}{\cellcolor[HTML]{FFFFFF}1.53}           & \multicolumn{1}{l|}{\cellcolor[HTML]{FFFFFF}0.73}          & \multicolumn{1}{l|}{\cellcolor[HTML]{FFFFFF}0.15}          & 0.88           & \multicolumn{1}{l|}{\cellcolor[HTML]{FFFFFF}\textbf{1.71}} & \multicolumn{1}{l|}{\cellcolor[HTML]{FFFFFF}\textbf{0.17}} & \multicolumn{1}{l|}{\cellcolor[HTML]{FFFFFF}\textbf{1.88}}  & \multicolumn{1}{l|}{\cellcolor[HTML]{FFFFFF}0.56}          & \multicolumn{1}{l|}{\cellcolor[HTML]{FFFFFF}0.11}          & 0.67           \\ \hline
\rowcolor[HTML]{EFEFEF} 
\multicolumn{1}{|l|}{\cellcolor[HTML]{EFEFEF}}                                    & SELECT                                                     & \multicolumn{1}{l|}{\cellcolor[HTML]{EFEFEF}1.50}          & \multicolumn{1}{l|}{\cellcolor[HTML]{EFEFEF}0.18}          & \multicolumn{1}{l|}{\cellcolor[HTML]{EFEFEF}1.68}           & \multicolumn{1}{l|}{\cellcolor[HTML]{EFEFEF}0.75}          & \multicolumn{1}{l|}{\cellcolor[HTML]{EFEFEF}0.14}          & 0.89           & \multicolumn{1}{l|}{\cellcolor[HTML]{EFEFEF}2.20}          & \multicolumn{1}{l|}{\cellcolor[HTML]{EFEFEF}0.18}          & \multicolumn{1}{l|}{\cellcolor[HTML]{EFEFEF}2.38}           & \multicolumn{1}{l|}{\cellcolor[HTML]{EFEFEF}0.55}          & \multicolumn{1}{l|}{\cellcolor[HTML]{EFEFEF}0.08}          & 0.63           \\ \cline{2-14} 
\rowcolor[HTML]{EFEFEF} 
\multicolumn{1}{|l|}{\cellcolor[HTML]{EFEFEF}}                                    & INSERT                                                     & \multicolumn{1}{l|}{\cellcolor[HTML]{EFEFEF}\textbf{1.01}} & \multicolumn{1}{l|}{\cellcolor[HTML]{EFEFEF}\textbf{0.18}} & \multicolumn{1}{l|}{\cellcolor[HTML]{EFEFEF}\textbf{1.19}}  & \multicolumn{1}{l|}{\cellcolor[HTML]{EFEFEF}0.67}          & \multicolumn{1}{l|}{\cellcolor[HTML]{EFEFEF}0.14}          & 0.81           & \multicolumn{1}{l|}{\cellcolor[HTML]{EFEFEF}\textbf{1.75}} & \multicolumn{1}{l|}{\cellcolor[HTML]{EFEFEF}\textbf{0.19}} & \multicolumn{1}{l|}{\cellcolor[HTML]{EFEFEF}\textbf{1.94}}  & \multicolumn{1}{l|}{\cellcolor[HTML]{EFEFEF}0.58}          & \multicolumn{1}{l|}{\cellcolor[HTML]{EFEFEF}0.08}          & 0.66           \\ \cline{2-14} 
\rowcolor[HTML]{EFEFEF} 
\multicolumn{1}{|l|}{\cellcolor[HTML]{EFEFEF}}                                    & UPDATE                                                     & \multicolumn{1}{l|}{\cellcolor[HTML]{EFEFEF}1.23}          & \multicolumn{1}{l|}{\cellcolor[HTML]{EFEFEF}0.18}          & \multicolumn{1}{l|}{\cellcolor[HTML]{EFEFEF}1.41}           & \multicolumn{1}{l|}{\cellcolor[HTML]{EFEFEF}0.77}          & \multicolumn{1}{l|}{\cellcolor[HTML]{EFEFEF}0.13}          & 0.90           & \multicolumn{1}{l|}{\cellcolor[HTML]{EFEFEF}2.29}          & \multicolumn{1}{l|}{\cellcolor[HTML]{EFEFEF}0.19}          & \multicolumn{1}{l|}{\cellcolor[HTML]{EFEFEF}2.48}           & \multicolumn{1}{l|}{\cellcolor[HTML]{EFEFEF}0.54}          & \multicolumn{1}{l|}{\cellcolor[HTML]{EFEFEF}0.08}          & 0.62           \\ \cline{2-14} 
\rowcolor[HTML]{EFEFEF} 
\multicolumn{1}{|l|}{\multirow{-4}{*}{\cellcolor[HTML]{EFEFEF}PostgreSQL}}        & DELETE                                                     & \multicolumn{1}{l|}{\cellcolor[HTML]{EFEFEF}1.26}          & \multicolumn{1}{l|}{\cellcolor[HTML]{EFEFEF}0.19}          & \multicolumn{1}{l|}{\cellcolor[HTML]{EFEFEF}1.45}           & \multicolumn{1}{l|}{\cellcolor[HTML]{EFEFEF}\textbf{0.54}} & \multicolumn{1}{l|}{\cellcolor[HTML]{EFEFEF}\textbf{0.13}} & \textbf{0.67}  & \multicolumn{1}{l|}{\cellcolor[HTML]{EFEFEF}1.84}          & \multicolumn{1}{l|}{\cellcolor[HTML]{EFEFEF}0.20}          & \multicolumn{1}{l|}{\cellcolor[HTML]{EFEFEF}2.04}           & \multicolumn{1}{l|}{\cellcolor[HTML]{EFEFEF}\textbf{0.49}} & \multicolumn{1}{l|}{\cellcolor[HTML]{EFEFEF}\textbf{0.08}} & \textbf{0.57}  \\ \hline
\rowcolor[HTML]{FFFFFF} 
\multicolumn{1}{|l|}{\cellcolor[HTML]{FFFFFF}}                                    & SELECT                                                     & \multicolumn{1}{l|}{\cellcolor[HTML]{FFFFFF}1.30}          & \multicolumn{1}{l|}{\cellcolor[HTML]{FFFFFF}0.19}          & \multicolumn{1}{l|}{\cellcolor[HTML]{FFFFFF}1.49}           & \multicolumn{1}{l|}{\cellcolor[HTML]{FFFFFF}\textbf{0.59}} & \multicolumn{1}{l|}{\cellcolor[HTML]{FFFFFF}\textbf{0.14}} & \textbf{0.73}  & \multicolumn{1}{l|}{\cellcolor[HTML]{FFFFFF}2.42}          & \multicolumn{1}{l|}{\cellcolor[HTML]{FFFFFF}0.19}          & \multicolumn{1}{l|}{\cellcolor[HTML]{FFFFFF}2.61}           & \multicolumn{1}{l|}{\cellcolor[HTML]{FFFFFF}0.65}          & \multicolumn{1}{l|}{\cellcolor[HTML]{FFFFFF}0.15}          & 0.80           \\ \cline{2-14} 
\rowcolor[HTML]{FFFFFF} 
\multicolumn{1}{|l|}{\cellcolor[HTML]{FFFFFF}}                                    & INSERT                                                     & \multicolumn{1}{l|}{\cellcolor[HTML]{FFFFFF}1.48}          & \multicolumn{1}{l|}{\cellcolor[HTML]{FFFFFF}0.20}          & \multicolumn{1}{l|}{\cellcolor[HTML]{FFFFFF}1.68}           & \multicolumn{1}{l|}{\cellcolor[HTML]{FFFFFF}0.69}          & \multicolumn{1}{l|}{\cellcolor[HTML]{FFFFFF}0.14}          & 0.83           & \multicolumn{1}{l|}{\cellcolor[HTML]{FFFFFF}2.21}          & \multicolumn{1}{l|}{\cellcolor[HTML]{FFFFFF}0.18}          & \multicolumn{1}{l|}{\cellcolor[HTML]{FFFFFF}2.39}           & \multicolumn{1}{l|}{\cellcolor[HTML]{FFFFFF}0.79}          & \multicolumn{1}{l|}{\cellcolor[HTML]{FFFFFF}0.15}          & 0.94           \\ \cline{2-14} 
\rowcolor[HTML]{FFFFFF} 
\multicolumn{1}{|l|}{\cellcolor[HTML]{FFFFFF}}                                    & UPDATE                                                     & \multicolumn{1}{l|}{\cellcolor[HTML]{FFFFFF}1.54}          & \multicolumn{1}{l|}{\cellcolor[HTML]{FFFFFF}0.19}          & \multicolumn{1}{l|}{\cellcolor[HTML]{FFFFFF}1.73}           & \multicolumn{1}{l|}{\cellcolor[HTML]{FFFFFF}0.69}          & \multicolumn{1}{l|}{\cellcolor[HTML]{FFFFFF}0.14}          & 0.83           & \multicolumn{1}{l|}{\cellcolor[HTML]{FFFFFF}2.19}          & \multicolumn{1}{l|}{\cellcolor[HTML]{FFFFFF}0.19}          & \multicolumn{1}{l|}{\cellcolor[HTML]{FFFFFF}2.38}           & \multicolumn{1}{l|}{\cellcolor[HTML]{FFFFFF}\textbf{0.47}} & \multicolumn{1}{l|}{\cellcolor[HTML]{FFFFFF}\textbf{0.14}} & \textbf{0.61}  \\ \cline{2-14} 
\rowcolor[HTML]{FFFFFF} 
\multicolumn{1}{|l|}{\multirow{-4}{*}{\cellcolor[HTML]{FFFFFF}Couchbase}}         & DELETE                                                     & \multicolumn{1}{l|}{\cellcolor[HTML]{FFFFFF}\textbf{1.18}} & \multicolumn{1}{l|}{\cellcolor[HTML]{FFFFFF}\textbf{0.20}} & \multicolumn{1}{l|}{\cellcolor[HTML]{FFFFFF}\textbf{1.38}}  & \multicolumn{1}{l|}{\cellcolor[HTML]{FFFFFF}0.73}          & \multicolumn{1}{l|}{\cellcolor[HTML]{FFFFFF}0.14}          & 0.87           & \multicolumn{1}{l|}{\cellcolor[HTML]{FFFFFF}\textbf{1.87}} & \multicolumn{1}{l|}{\cellcolor[HTML]{FFFFFF}\textbf{0.18}} & \multicolumn{1}{l|}{\cellcolor[HTML]{FFFFFF}\textbf{2.05}}  & \multicolumn{1}{l|}{\cellcolor[HTML]{FFFFFF}0.72}          & \multicolumn{1}{l|}{\cellcolor[HTML]{FFFFFF}0.15}          & 0.87           \\ \hline
\end{tabular}
\end{table*}

\section{Experiment}
We initiated the development of DBJoules under the presumption that there might be variations in energy consumption values across different databases. To confirm this hypothesis, we conducted an empirical experiment to assess the energy consumption levels of these four databases when executing fundamental queries, namely SELECT, INSERT, DELETE, and UPDATE, utilizing \textit{DBJoules}. For this experiment, we utilized the Netflix Userbase dataset\footnote{\url{https://www.kaggle.com/datasets/arnavsmayan/netflix-userbase-dataset}} and and SMS Spam collection dataset\footnote{\url{https://archive.ics.uci.edu/dataset/228/sms+spam+collection}}. Our aim with this empirical analysis to identify whether actually there are differences in the energy consumption values of queries across various databases. The experiment is performed on two distinct systems equipped with Intel and AMD processor respectively and having following configurations:
\begin{itemize}
    \item \textbf{System1:} AMD(R) Ryzen 5 3500U CPU with a frequency of 2.10GHz, 4 cores, and 8GB RAM
    \item \textbf{System2:} Intel(R) Core(TM) i5-1135G7 CPU with a frequency of 2.40GHz, 4 cores, and 8GB RAM
\end{itemize}

To minimize the influence of noise, we manually terminated all background processes and executed a script to halt other concurrent background processes. Each query was executed ten times on all four databases to ensure precise energy consumption measurements, and the average energy consumption was recorded, following the methodology outlined by Georgiou et al. \cite{georgiou2022green}. However, some factors such as unnoticed background processes could add noise in our measurements. Therefore, to reduce the impact of potential noise we chose to execute these queries by shuffling at random rather than simply executing it sequentially. We maintained the system in an idle state for 30 seconds after the execution of each query to ensure system stability and mitigate the impact of power tail states \cite{bornholt2012model}. Energy readings obtained from the \textit{DBJoules} tool were saved in a \textit{DBJoules\_output.csv} file\footnote{Evaluation scripts and results for energy consumption values for each trial can be accessed at \url{https://github.com/rishalab/DBJoules/tree/main/Experiments}} after each iteration, including the execution time for each query and the corresponding energy consumption readings for CPU and RAM in Joules. 

The mean energy consumption values for the database queries executed on both systems are presented in Table \ref{tab:my-table}. Our findings indicate that the Intel system exhibits a minimum 50\% reduction in energy consumption across all queries and databases, compared to the AMD processor system. Moreover, CPU energy consumption surpasses that of RAM, due to the CPU's role in managing the most of database operations. Among the databases, MongoDB demonstrates overall lower energy consumption for SELECT, UPDATE, and DELETE queries, while PostgreSQL better in handling INSERT queries. Notably, the variations in energy consumption values for specific queries across all four databases fall within the range of 7\% to 38\%. However, we emphasize that more rigorous experiments are necessary to confirm these findings, and \textit{DBJoules} represents our initial step towards the development of a standardized tool capable of measuring database energy consumption.

\section{Discussion and Future work}
The current version of DBJoules exhibits certain limitations that should be addressed. Firstly, it is supported only on the Windows operating system (OS). Our future plans involve extending its support to Linux and macOS. Secondly,the tool currently supports only four basic queries: SELECT, INSERT, UPDATE, and DELETE. While these are fundamental queries, they serve to demonstrate the significant variation in energy consumption across databases, as shown in Table \ref{tab:my-table}. However, we aim to include more complex queries in the future. 




The values provided by the psutil package to DBJoules include system-wide CPU utilization and memory allocation which can be influenced by noise interference and concurrent background processes. Therefore, users of this tool are advised to manually terminate any background processes. Furthermore, the calculation of CPU energy consumption relies on TDP values obtained from the dataset in \cite{budennyy2022eco2ai} and are not real-time and may not accurately reflect the current state of the host machine. Consequently, there may be a margin of error in calculating CPU energy consumption. However, this bias is consistent across all databases, reducing its impact on error. It's important to note that DBJoules is currently limited to systems equipped with Intel and AMD processors.


To address challenges in measuring energy consumption within software systems due to voltage spikes and daemon threads, users should obtain mean energy consumption values by running the code snippet multiple times during empirical experiments. Additionally, the phenomenon of continued high power draw after task completion, known as the power tail state \cite{bornholt2012model}, necessitates briefly idling the host machine before tool execution to mitigate its impact.


In future directions, it's important to consider GPU-accelerated database management systems \cite{shanbhag2020study}. This includes the need to assess energy consumption from the GPU perspective in addition to CPU and RAM utilization, as currently considered in \textit{DBJoules}.

\section{Conclusion}
\textit{DBJoules} is a tool which measures the energy consumption of four popular databases. This tool can serve as a valuable resource for developers and practitioners seeking to develop energy-aware applications using databases. It represents one of the initial steps in investigating the energy consumption of database queries. Through an evaluation of this tool on two systems and a comparison of energy consumption across four databases, we have demonstrated its effectiveness in providing meaningful results.

\bibliographystyle{ieeetr}
\bibliography{reference}

\end{document}